\documentclass{osa-article}

\journal{oe}

\articletype{Research Article}

\usepackage{lineno}

\begin{document}

\title{The Faraday effect in magnetoplasmonic nanostructures with spatial modulation of magnetization}

\author{O.V. Borovkova,\authormark{1,2,*} S.V. Lutsenko,\authormark{1,2} D.A. Sylgacheva,\authormark{1,2} A.N. Kalish,\authormark{1,2,3} and V.I. Belotelov\authormark{1,2}}

\address{\authormark{1}Russian Quantum Center, Novaya str. 100, Skolkovo, Moscow Region 143025, Russia\\
\authormark{2}Faculty of Physics, Lomonosov Moscow State University, Leninskie Gory, Moscow 119991, Russia\\
\authormark{3}NTI Center for Quantum Communications, National University of Science and Technology MISiS, Leninsky Prospekt 4, 119049 Moscow, Russia}

\email{\authormark{*}o.borovkova@rqc.ru} 



\begin{abstract}
The magneto-optical Faraday effect in the magnetoplasmonic nanostructures with nonuniform, periodically modulated spatial distribution of the magnetization is considered. It is shown that in such nanostructures the Faraday effect can experience the resonant enhancement in the spectral range of the plasmonic and waveguide optical modes excitation. It happens both for $s$ and $p$-polarized light. Such an effect can serve for the spectrally selective detection of the short spin waves in the magnetic materials.
\end{abstract}

\section{Introduction}

An engineering of the optical properties of artificial materials and nanostructures enables their application in various and diverse areas. Optical characteristics can be managed by the nanostructuring \cite{MaierPlasmonics, Stockman:2018, Song:2019} or by controlling of the dielectric properties via illumination by powerful optical beam, short optical pulses, heating, application of the external magnetic field, etc. The latter case is relevant when a nanostructure contains magnetic components \cite{Park:2003, Wurtz:2008, Belotelov:2011, Baryshev:2013, Armelles:2013, Atoneche:2010, Maccaferri:2016, Lutsenko:2021, Kekesi:2015, Borovkova:2019, Kreilkamp:2013, Floess:2018, Chekhov:2014, Lukyanchuk:2010, Kuzmichev:2020, Krichevsky:2020}, and is the most interesting in context of the present work.

Depending on the direction of the applied magnetic field the dielectric tensor of a material acquires different nondiagonal components. As a result the optical properties of a magnetic material change that leads to the modification of the reflectance and transmittance of the structure. The mutual arrangement of the magnetic field and input light plane as well as the polarization state of light determine the certain type of the observed magneto-optical effect \cite{MOBook}.

The magneto-optical Faraday effect emerges when the magnetic field is directed along the light wavevector. Such configuration is called polar. The linearly polarized input light reveals the polarization rotation after the transmittance through the magnetic material. The Faraday effect has been thoroughly studied in separate magnetic layers \cite{MOBook} as well as in the magnetoplasmonic nanostructures \cite{Kushwaha:1987, Lukyanchuk:2010, Belotelov:2014, Levy:19, Kuzmichev:2020, Krichevsky:2020}. It was shown that in the magnetoplasmonic nanostructures the Faraday effect experiences the resonant peculiarities in the spectral range of the surface plasmon polariton (SPP) and waveguide (WG) optical modes excitation \cite{Kushwaha:1987}. In presence of magnetization of the dielectric material, the SPP mode obtains the $s$-polarized component of the mode that depends on the material magnetization \cite{kalish2014transformation}. Similarly, the hybrid WG modes in the ferrimagnetic layer covered by metal contain the $s$-polarized component \cite{Kushwaha:1987, Belotelov:2014, kalish2014transformation}. It should be noted that as soon as the nanostructure becomes demagnetized the $s$-polarized components of optical modes vanish. 

Until now the magneto-optical effect in polar configuration of the magnetic field has been addressed for uniformly magnetized samples. The magnetization was supposed to be constant in the plane of the sample. However, the spatial modulation of the magnetization opens up new opportunities for the control of magneto-optical effects. For instance, it was shown in recent work \cite{BorovkovaMDPI:2022} that the spatial modulation of the magnetization allows to control the magneto-optical effect in the transverse configuration. Interestingly, a similar effect can be observed in magnetoplasmonic structures with spatial symmetry breaking of the plasmonic grating \cite{Borovkova:2020, Borovkova:2022}.

Spatial magnetization modulation in a ferrimagnetic layer can be created in by different ways. It could be induced by internal electromagnetic fields due to the alternating magnetic domains \cite{Hubert:2008, Barturen:2012}, or it can be generated by external impacts like by compression and stretching \cite{Dho:2003, Dai:2019} or by interaction of ferroelectric and ferromagnetic layers in heterostructures \cite{Lahtinen:2011}. Apart from that, the spin waves create the periodic modulation of the magnetization in ferrimagnetic layers \cite{Satoh:2012, Nikitov:2015, Kimel:2005, Kirilyuk:2010, Kozhaev:2018, Savochkin:2017, BorovkovaMDPI:2022}. 

In this paper we are interested in the properties of the magneto-optical Faraday effect in the magnetoplasmonic nanostructure with periodic spatial modulation of the magnetization in the ferrimagnetic layer. A combination of nonuniform distribution of the magnetic properties of the ferrimagnetic material and nonuniform optical fields due to the excitation of plasmonic and/or waveguide optical modes originates resonant enhancement of the Faraday effect in the spectral range of optical modes. The effect is considered for both $p$- and $s$-polarized input light. Further these results can be explored for the purposes of short spin waves detection.

\section{The addressed magnetoplasmonic nanostructure}

\begin{figure}[ht]
\centering
\includegraphics[width=0.9\linewidth]{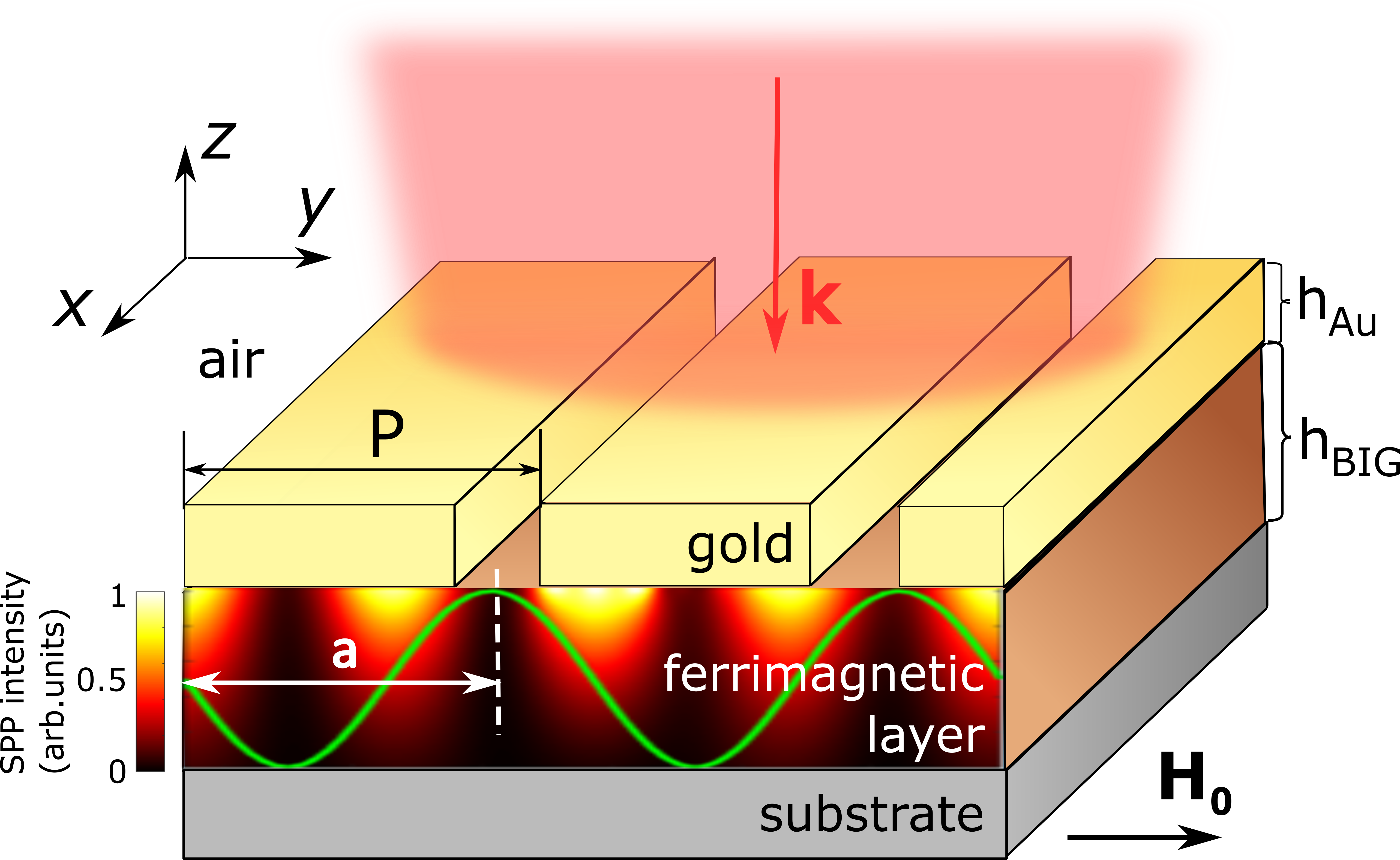}
\caption{The scheme of the analyzed magnetoplasmonic nanostructure with spatial modulation of the magnetization in the ferrimagnetic layer. To address the Faraday effect the structure is illuminated by linearly $s$- or $p$-polarized light with the wavevector normal to the surface of the sample. An example of the spatial distribution of the electromagnetic field in ferrimagnetic layer is shown by yellow-red-black colormap. It corresponds to the excitation of the SPP optical modes at the [metal]/[ferrimagnetic layer] interface.}
\label{fig:scheme}
\end{figure}

The addressed magnetoplasmonic nanostructure is given in Figure \ref{fig:scheme}. The ferrimagnetic layer of bismuth-substituted yttrium iron garnet (BIG) (its magnetic properties can be found, e.g., in \cite{Popova:2012, Deb:2012}) of thickness $h_{BIG}=100$~nm is on top of the nonmagnetic substrate of gadolinium gallium garnet (GGG). The ferrimagnetic layer is covered by the subwavelength gold grating of thickness $h_{Au}=80$~nm. The period is varied to explore the magneto-optical properties. The air gap between the gold stripes is varied. The external magnetic field can be applied along $y$-axis or along $z$-axis (not shown). In both cases the magnetization modulation created in the magnetic layer due to the excitation of spin waves will have the $m_x$ component (for the details see \cite{BorovkovaMDPI:2022}). In Fig.~\ref{fig:scheme} the magnetization modulation is shown by green line. To address the Faraday effect the magnetoplasmonic nanostructure is illuminated from top by linearly $s$- or $p$-polarized light with the wavevector normal to the surface of the sample.

We are interested in how the variation of the magnetization modulation, namely, its period, influences the resulting Faraday effect. We address the scheme given in Fig.~\ref{fig:scheme}, when the magnetization modulation direction coincides with the direction of modulation of a one-dimensional plasmonic grating. We describe  magnetization modulation as a modulation of the local gyration of the ferrimagnetic material

\begin{equation}
g(y) = g_0 \sin(k_{SW}y+k_{SW}a),
\label{eq:gyration}
\end{equation}
here $g$ is related to magnetization-induced nondiagonal components of the dielectric tensor ($g=i\varepsilon_{yz}=-i\varepsilon_{zy}$), $g_0$ is the amplitude of the magnetization modulation, $k_{SW}$ is the wavevector of the corresponding spin wave that can induce such modulation, $k_{SW}=2\pi/\lambda_{SW}$, and $a$ is a shift between `zero' of the gyration modulation and the center of the air gap between gold stripes producing the phase shift between them. Parameter $a$ can take any values in the interval $[0,P)$, where $P$ is a period of the plasmonic grating. 

Let's introduce the parameter $k$ that reflects the ratio between the period of the plasmonic grating, $P$, and the magnetization modulation period. In Fig. \ref{fig:scheme} the period of the plasmonic grating and the magnetization modulation are equal, so $k=1$. To simplify the description we will use $k=0$ to refer to the case of the uniform magnetization of the ferrimagnetic layer, i.e. the case of absent modulation. Further, the smaller is the magnetization modulation period, the greater is parameter $k$.

The described nanostructure is analyzed by numerical simulation of the periodic magnetoplasmonic nanostructures by the rigorous coupled-wave analysis (RCWA) \cite{Moharam:1995, Li:2003}. The RCWA method provides calculation of both far-field characteristics (such as the intensity and polarization of scattered waves for all diffraction orders) and near-field distribution of electromagnetic field components.

\section{The Faraday effect in the magnetoplasmonic nanostructure with spatially modulated magnetization for $p$-polarized input light}

The Faraday effect (FE) reveals itself as a rotation of the polarization plane of light (the Faraday angle, $\Psi$) when it passes through the magnetic material or nanostructure. As it was shown in \cite{Belotelov:2014} the FE experiences the resonant features when the conditions for optical modes excitation are satisfied. Here we are interested in how the FE varies with the modulation of the magnetization in the ferrimagnetic layer.

\begin{figure}[ht]
\begin{minipage}[h]{0.45\linewidth}
\center{\includegraphics[width=\linewidth]{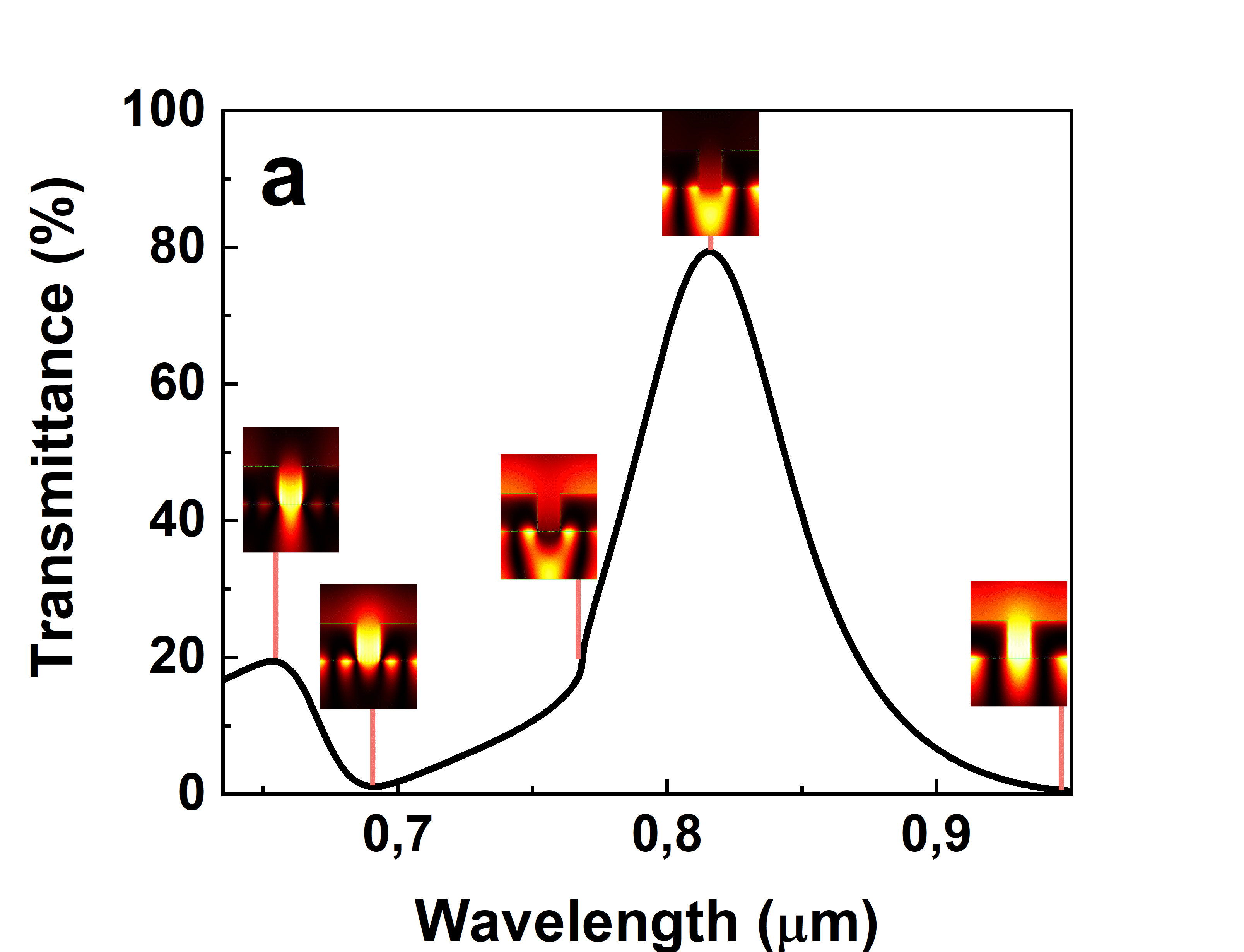}}
\end{minipage}
\hfill
\begin{minipage}[h]{0.45\linewidth}
\center{\includegraphics[width=\linewidth]{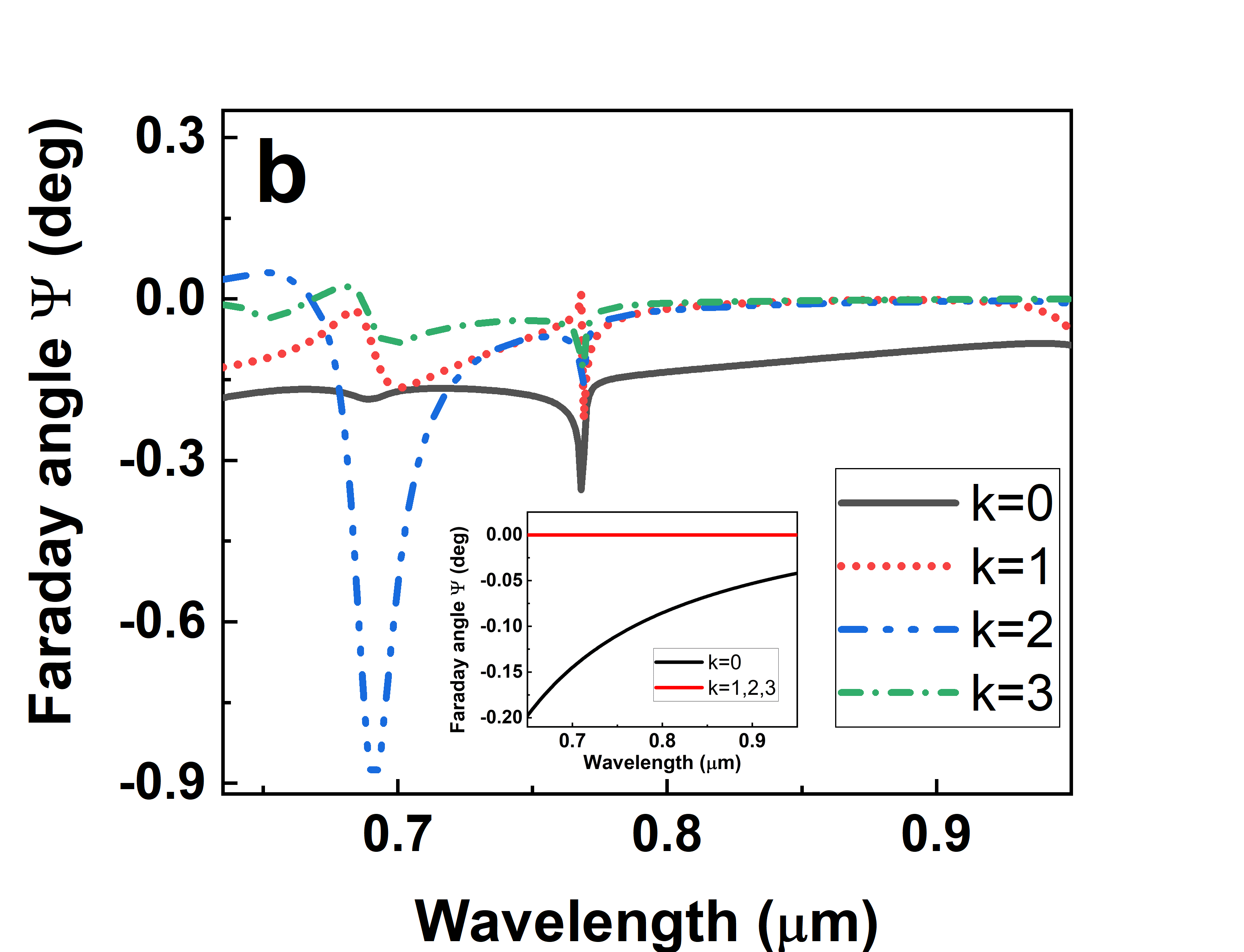}}
\end{minipage} 
\caption{a) Transmittance spectrum of the considered magnetoplasmonic nanostructure. The insets illustrate the spatial distribution of the electromagnetic field inside the ferrimagnetic layer inside one period of the plasmonic grating. b) The Faraday effect spectrum versus the wavelength of the incident light for various values of the parameter $k$. In both plots the period of the gold grating is 390~nm, the air gap width is 85~nm. Input light is $p$-polarized. The inset shows the comparison of the FE spectra for uniform and spatially modulated magnetization for ferrimagnetic layer without the plasmonic grating.}
\label{fig:FE_TM}
\end{figure}

In Fig. \ref{fig:FE_TM}a the transmittance spectrum of the considered magnetoplasmonic nanostructure with the period of the gold grating is 390~nm and the air gap width is 85~nm is shown. It is illuminated by $p$-polarized light. The shift between the plasmonic grating and the magnetization modulation is taken as $a=42.5$~nm. As one can see the SPP mode is excited at wavelength 689~nm. The corresponding spectrum of the Faraday angle $\Psi$ is given in Fig. \ref{fig:FE_TM}b. The solid line in the plot refers to the magnetoplasmonic nanostructure with uniform magnetization of the ferrimagnetic layer. One can see that this spectrum has slight peculiarity related to the SPP mode, but its magnitude is too small. More pronounced is the peculiarity of Faraday angle is at the wavelength of 768~nm that corresponds to the hybrid plasmon-waveguide mode.

However, as soon as the magnetization of ferrimagnetic layer is spatially modulated, i.e. $k=1,2,...$, the peculiarities of the spectrum of the Faraday angle increase. It happens for both SPP modes and hybrid SPP-waveguide modes. But the greatest resonant enhancement of the FE up to $0.9^\circ$ is observed for the SPP mode when $k=2$, that means that the period of the plasmonic grating is equal to two periods of the magnetization modulation. This effect can be explained by the fact that the SPP wavelength is close to the period the magnetization modulation when $k=2$, as can be seen from the SPP field distribution shown in Fig. \ref{fig:FE_TM}a. As a result the effectiveness of the FE resonantly increases.

To prove the key importance of the SPP modes for the observed resonant enhancement of the FE we calculated the spectra of the Faraday rotation in the similar ferrimagnetic layer without plasmonic grating on top. The resulting spectra are given in the inset of Fig. \ref{fig:FE_TM}b. The black solid curve shows the smooth spectrum of the FE in uniformly magnetized ferrimagnetic layer. The conditions for the optical modes excitation are not satisfied, so there are no peculiarities in the FE spectrum. On the contrary, for three different spatial periods of the magnetization modulation, the observed Faraday rotation is equal to zero. The optical light beam has a diameter much greater than the period of the magnetization modulation, so averaged impact on the effect is zero.

\begin{figure}[ht]
\centering
\includegraphics[width=0.7\linewidth]{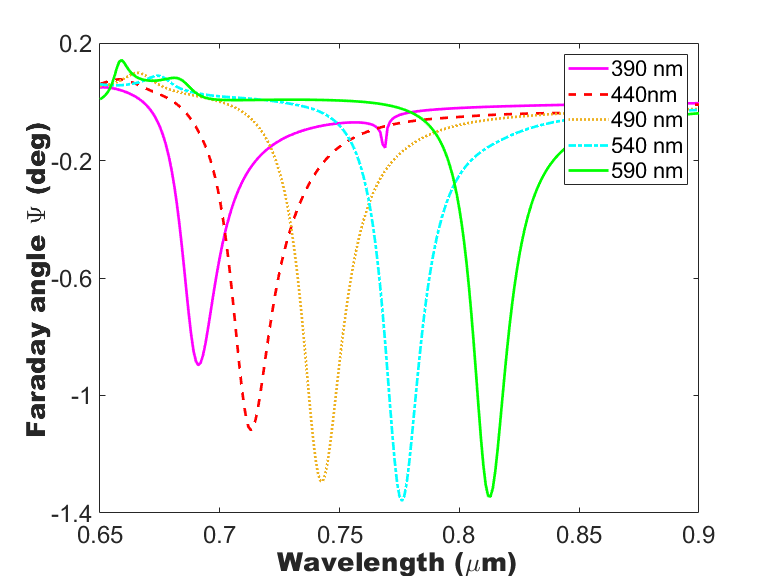}
\caption{The Faraday effect spectrum versus the wavelength of the incident light for different periods of the gold grating for $k=2$. Input light is $p$-polarized.}
\label{fig:FEvarP}
\end{figure}

As soon as the observed FE is SPP-assisted it should depend on the parameters of the plasmonic grating, namely, its period and the air gap. Although the air gap variation along with the fixed period of the grating doesn't have any significant effect on the properties of the observed magneto-optical response. At the same time the increase of the plasmonic period $P$ the spectral position of the SPP resonance experiences a red shift. Note that the other parameters are fixed, except the period of the magnetization modulation that is equal to the half of the corresponding plasmonic grating period $P$, so the ratio $k=2$. As one can see from Fig.~\ref{fig:FEvarP} this red shift is accompanied by the gradual increase of the magnitude of the Faraday rotation $\Psi$. The most pronounced effect is observed when the plasmonic period of the grating is equal to 540~nm (turquoise dot-dash line). In this case the peak value of the Faraday rotation is $1.5$ times greater than for plasmonic grating with period 390~nm. This optimization is suitable when the spectral position of the resonance isn't crucial. On the contrary, even for the fixed frequency in the shown spectral range the magnitudes of the Faraday rotation are measurable and detectable.

\begin{figure}[ht]
\begin{minipage}[h]{0.45\linewidth}
\center{\includegraphics[width=\linewidth]{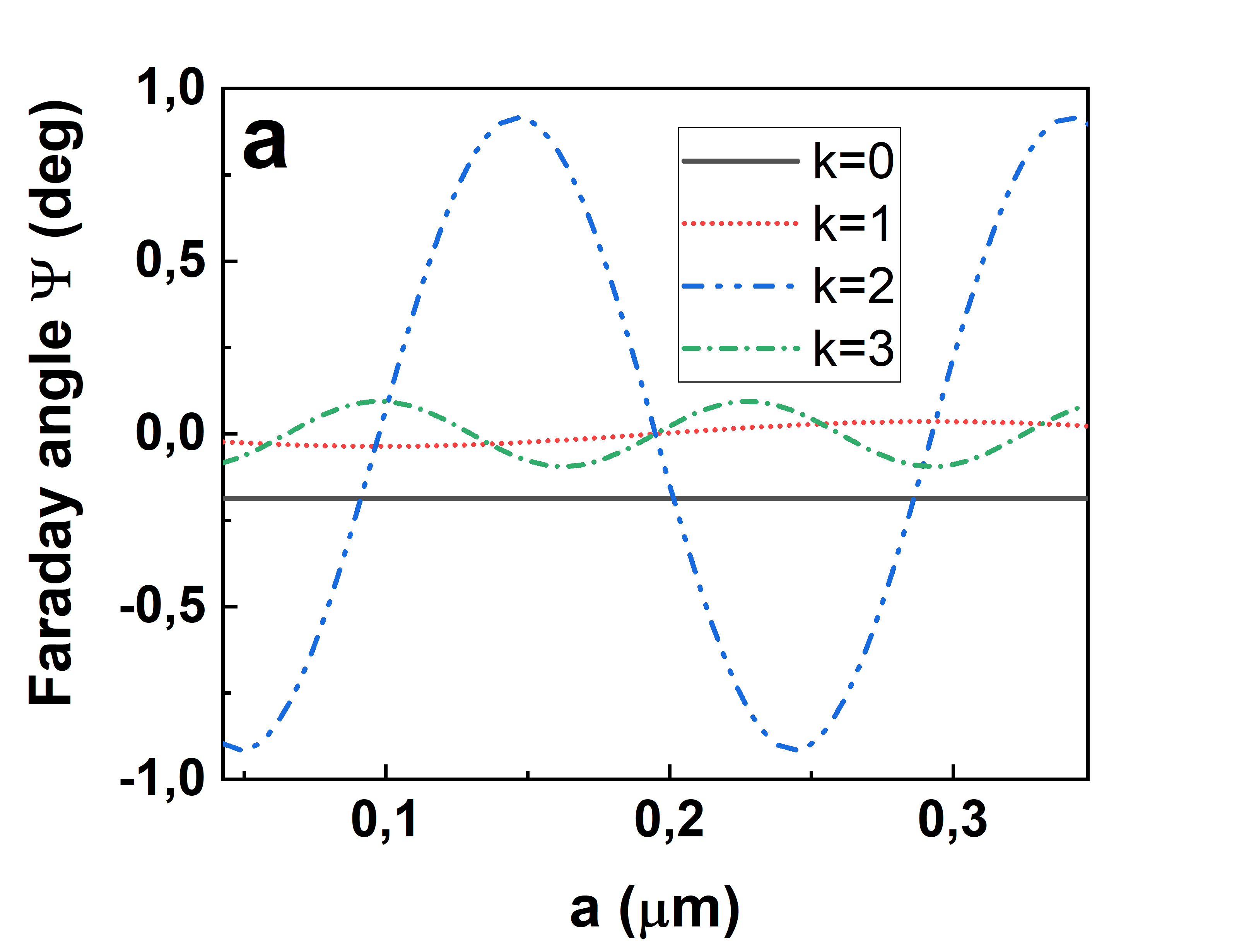}}
\end{minipage}
\hfill
\begin{minipage}[h]{0.45\linewidth}
\center{\includegraphics[width=\linewidth]{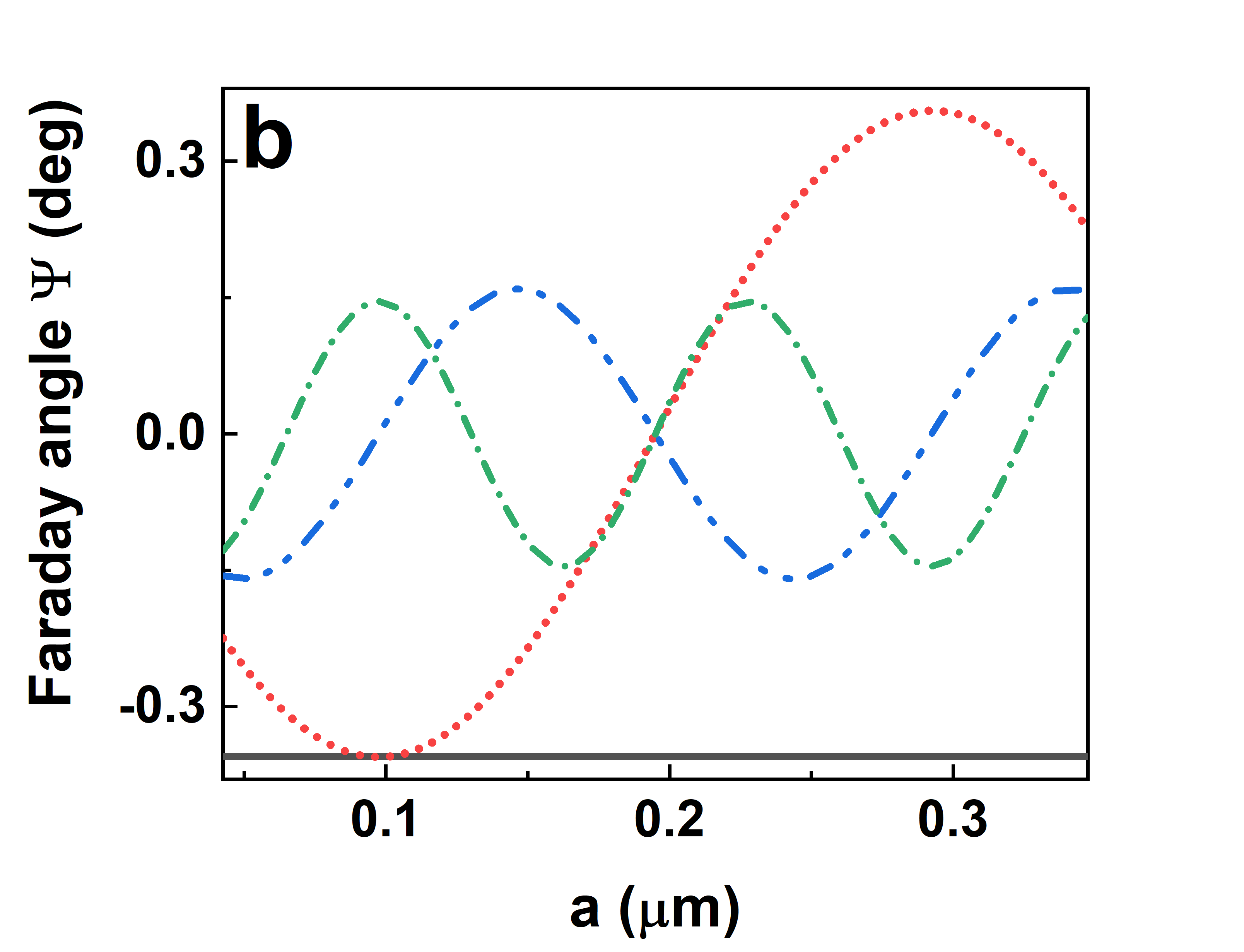}}
\end{minipage} 
\caption{The Faraday rotation versus the shift between the plasmonic grating and the magnetization modulation for different values of parameter $k$ (solid black lines refer to the uniform magnetization of the ferrimagnetic layer). The Faraday angle is taken at the wavelength of a) 689~nm, b) 768~nm. In both plots the period of the gold grating is 390~nm, the air gap width is 85~nm. Input light is $p$-polarized.}
\label{fig:FEvara}
\end{figure}

The resulting Faraday rotation angle $\Psi$ depends on the shift $a$ between the plasmonic grating and the magnetization modulation. It is shown in Fig.~\ref{fig:FEvara} how the magneto-optical effect depends on the shift $a$ for two different wavelength, 689~nm and 768~nm. Both wavelengths correspond to the peculiarities in the spectra of the FE due to the excitation of the optical modes (see Fig.~\ref{fig:FE_TM}).

The results in Fig.~\ref{fig:FEvara} are mostly interesting from the point of view of spin waves detection. Magnetization modulation appearing in case of the spin waves propagation will dynamically move along $y$-axis (see Fig.~\ref{fig:scheme}) and the FE measured dynamically will demonstrate the oscillations. The period of such oscillations corresponds to the period of the magnetization modulation that facilitates the spectrally selective detection of the propagating spin waves. The similar mechanism was reported earlier in \cite{BorovkovaMDPI:2022}.

\section{The Faraday effect in the magnetoplasmonic nanostructure with spatially modulated magnetization for $s$-polarized input light}

It is interesting to analyze how the magnetization modulation affects the transmittance and the FE if the input light is $s$-polarized. In this case the SPP modes are not excited, but the waveguide modes in the ferrimagnetic layer provide the conditions for the peculiarities occuring in the spectrum of the FE. The $s$-polarized input light in such scheme facilitates the measurements because the transmission coefficient in the spectral range of the resonances can be even greater than for $p$-polarized light.

\begin{figure}[ht]
\centering
\includegraphics[width=0.6\linewidth]{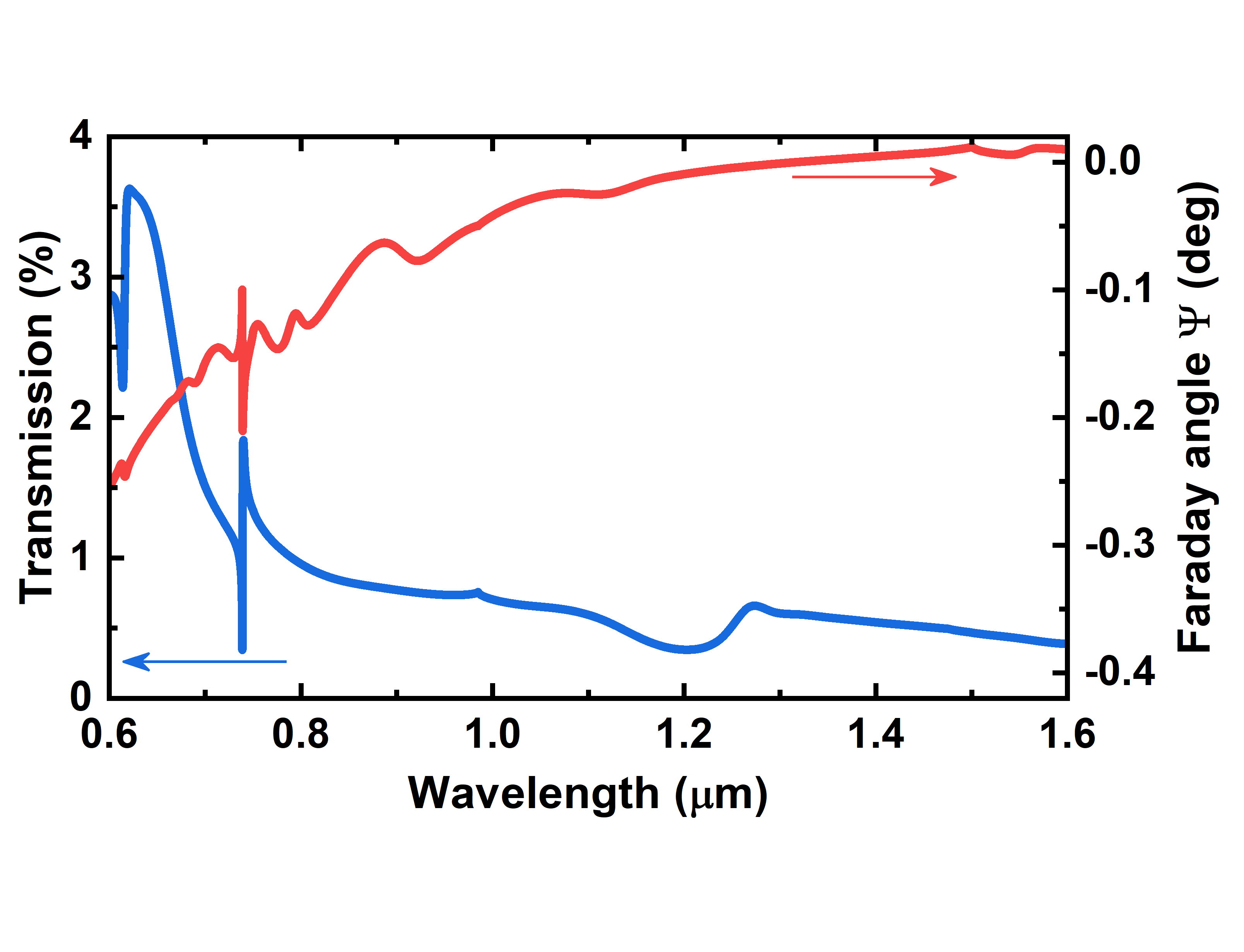}
\caption{Transmittance (blue line) and Faraday effect (red line) versus spectra for $s$-polarized input light. The plasmonic grating period is 1500~nm. No magnetization modulation.}
\label{fig:TE_FE}
\end{figure}

The transmittance (left axis, blue line) and the FE (right axis, red line) spectra in the optimized magnetoplasmonic nanostructure with uniform magnetization of the ferrimagnetic layer are given in Fig.~\ref{fig:TE_FE}. The plasmonic grating period is 1500~nm, the air gap is 150~nm. One can see the small peculiarities in the spectrum of the FE related to the excitation of the waveguide optical modes at 614~nm and 739~nm.

\begin{figure}[ht]
\centering
\includegraphics[width=0.9\linewidth]{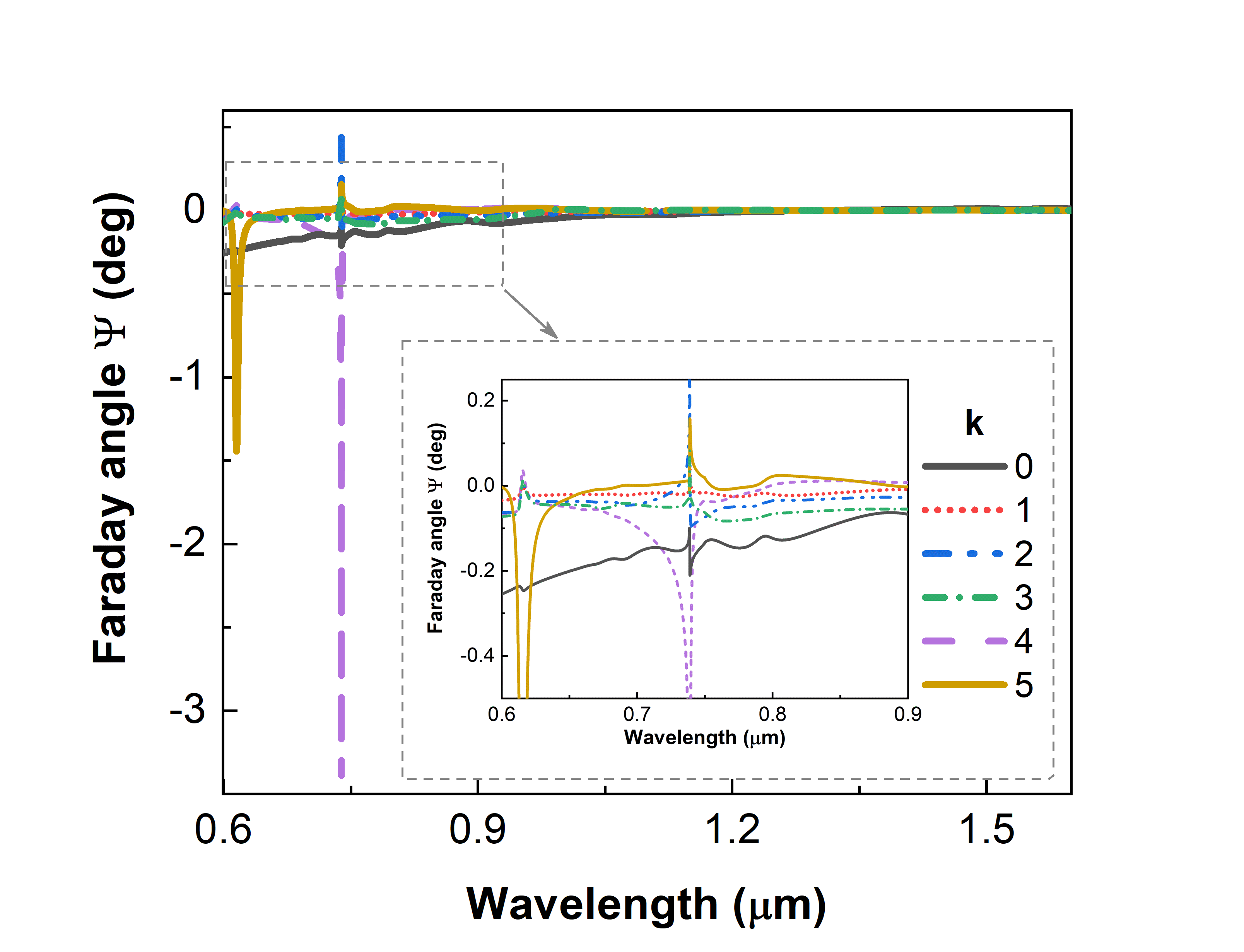}
\caption{Spectrum of the Faraday effect for various periods of the magnetization modulation (i.e., different values of the parameter k) for $s$-polarized input light. The plasmonic grating period is 1500~nm. The inset provides the zoomed picture of the Faraday rotation dependencies in the spectral range $0.6-1.1~\mu m$.}
\label{fig:TE_FE_vark}
\end{figure}

It was revealed that the resonant enhancement of the magneto-optical effect for $s$-polarized light can be observed for high values of the magnetization modulation. In the Fig.~\ref{fig:TE_FE_vark} the Faraday rotation spectra for different periods of the magnetization modulation (determined by parameter $k$) are given. The greatest resonances of the magneto-optical effect are supported by modulation periods of 375~nm (corresponding to $k=4$) and 300~nm (i.e., $k=5$). These periods coincide with the effective wavelength of the excited waveguide modes in the ferrimagnetic layer and therefore provide the resonant enhancement of the Faraday rotation.

So, the $s$-polarized light is preferable in case of ultra-short spin waves detection. 

\section{Conclusions}

To sum up, for the first time to our knowledge the magneto-optical Faraday effect was addressed in the magnetoplasmonic nanostructures with periodic modulation of the magnetization. It was revealed that the nonuniform distribution of the gyration in the ferrimagnetic material combined with the excitation of the plasmonic and waveguide optical modes provide the significant resonant enhancement of the Faraday rotation when the effective wavelength of the optical mode coincides with the period of the magnetization modulation. The effect was shown for both $p$- and $s$-polarized input light. The reported effect can be also employed for the spectrally selective detection of the spin waves in the ferrimagnetic component of the nanostructure. With the propagation of the spin waves and the corresponding spatial shift of the magnetization modulation, the dynamically measured Faraday effect oscillates with the period of the spin waves. 

\begin{backmatter}
\bmsection{Funding}
This study was supported by Russian Science Foundation (project no. 20-72-10159).

\bmsection{Acknowledgments}
The authors thank M.A. Kozhaev for valuable discussions. O.V.B. thanks the personal support by the Foundation for the Advancement of Theoretical Physics and Mathematics ''BASIS''. A.N.K. and V.I.B. are members of the Interdisciplinary Scientific and Educational School of Moscow University “Photonic and Quantum technologies. Digital medicine”.

\bmsection{Disclosures}
The authors declare no conflicts of interest.

\bmsection{Data Availability Statement}
Data underlying the results presented in this paper are not publicly available at this time but may be obtained from the authors upon reasonable request.


\end{backmatter}

\bibliography{main}

\begin{thebibliography}{10}
\newcommand{\enquote}[1]{``#1''}

\bibitem{MaierPlasmonics}
S.~Maier, \emph{Plasmonics: Fundamentals and Applications} (Springer New York,
  NY, 2007).

\bibitem{Stockman:2018}
M.~I. Stockman, K.~Kneipp, S.~I. Bozhevolnyi, S.~Saha, and A.~D. et~al,
  \enquote{Roadmap on plasmonics,} {\protect\JournalTitle{Journal of Optics}}
  \textbf{20}, 043001 (2018).

\bibitem{Song:2019}
M.~Song, D.~Wang, S.~Peana, S.~Choudhury, P.~Nyga, Z.~A. Kudyshev, H.~Yu,
  A.~Boltasseva, V.~M. Shalaev, and A.~V. Kildishev, \enquote{Colors with
  plasmonic nanostructures: A full-spectrum review,}
  {\protect\JournalTitle{Applied Physics Reviews}} \textbf{6}, 041308 (2019).

\bibitem{Park:2003}
J.~Park, M.~Inoue, J.~Cho, K.~Nishimura, and H.~Uchida, \enquote{An optical
  micro-magnetic device: magnetic-spatial light modulator,}
  {\protect\JournalTitle{J. Magn.}} \textbf{8}, 50 (2003).

\bibitem{Wurtz:2008}
G.~Wurtz, W.~Hendren, R.~Pollard, R.~Atkinson, L.~Guyader, A.~Kirilyuk,
  T.~Rasing, I.~Smolyaninov, and A.~Zayats, \enquote{Controlling optical
  transmission through magneto-plasmonic crystals with an external magnetic
  field,} {\protect\JournalTitle{Adv. Opt. Mater.}} \textbf{10}, 105012 (2008).

\bibitem{Belotelov:2011}
V.~Belotelov, I.~Akimov, M.~Pohl, V.~Kotov, S.~Kasture, A.~Vengurlekar, A.~V.
  Gopal, D.~Yakovlev, A.~Zvezdin, and M.~Bayer, \enquote{Enhanced
  magneto-optical effects in magnetoplasmonic crystals,}
  {\protect\JournalTitle{Nature Nanotechnology}} \textbf{6}, 370--376 (2011).

\bibitem{Baryshev:2013}
A.~Baryshev, H.~Uchida, and M.~Inoue, \enquote{Peculiarities of
  plasmon-modified magneto-optical response of gold-garnet structures,}
  {\protect\JournalTitle{J. Opt. Soc. Am. B}} \textbf{30}, 2371–--2376
  (2013).

\bibitem{Armelles:2013}
G.~Armelles, A.~Cebollada, A.~García-Martín, and M.~González,
  \enquote{Magnetoplasmonics: Combining magnetic and plasmonic
  functionalities,} {\protect\JournalTitle{Adv. Opt. Mater.}} \textbf{1},
  10–35 (2013).

\bibitem{Atoneche:2010}
F.~Atoneche, A.~Kalashnikova, A.~Kimel, A.~Stupakiewicz, A.~Maziewski,
  A.~Kirilyuk, and T.~Rasing, \enquote{Large ultrafast photoinduced magnetic
  anisotropy in a cobalt-substituted yttrium iron garnet,}
  {\protect\JournalTitle{Phys. Rev. B}} \textbf{81}, 214440 (2010).

\bibitem{Maccaferri:2016}
N.~Maccaferri, L.~Bergamini, M.~Pancaldi, M.~K. Schmidt, M.~Kataja, S.~van
  Dijken, N.~Zabala, J.~Aizpurua, and P.~Vavassori, \enquote{Anisotropic
  nanoantenna-based magnetoplasmonic crystals for highly enhanced and tunable
  magneto-optical activity,} {\protect\JournalTitle{Nano Lett.}} \textbf{16},
  2533 (2016).

\bibitem{Lutsenko:2021}
S.~V. Lutsenko, M.~A. Kozhaev, O.~V. Borovkova, A.~N. Kalish, A.~G. Temiryazev,
  S.~A. Dagesyan, V.~N. Berzhansky, A.~N. Shaposhnikov, A.~N. Kuzmichev, and
  V.~I. Belotelov, \enquote{Multiperiodic magnetoplasmonic gratings fabricated
  by the pulse force nanolithography,} {\protect\JournalTitle{Opt. Lett.}}
  \textbf{46}, 4148--4151 (2021).

\bibitem{Kekesi:2015}
R.~Kekesi, D.~Mart\'{i}n-Becerra, D.~Meneses-Rodr\'{i}guez,
  F.~Garc\'{i}a-P\'{e}rez, A.~Cebollada, and G.~Armelles, \enquote{Enhanced
  nonreciprocal effects in magnetoplasmonic systems supporting simultaneously
  localized and propagating plasmons,} {\protect\JournalTitle{Opt. Express}}
  \textbf{23}, 8128--8133 (2015).

\bibitem{Borovkova:2019}
O.~Borovkova, F.~Spitzer, V.~Belotelov, I.~Akimov, A.~Poddubny, G.~Karczewski,
  M.~Wiater, T.~Wojtowicz, A.~Zvezdin, D.~Yakovlev, and M.~Bayer,
  \enquote{Transverse magneto-optical kerr effect at narrow optical
  resonances,} {\protect\JournalTitle{Nanophotonics}} \textbf{8}, 287--296
  (2019).

\bibitem{Kreilkamp:2013}
L.~Kreilkamp, V.~Belotelov, J.~Chin, S.~Neutzner, D.~Dregely, T.~Wehlus,
  I.~Akimov, M.~Bayer, B.~Stritzker, and H.~Giessen, \enquote{Waveguide-plasmon
  polaritons enhance transverse magneto-optical kerr effect,}
  {\protect\JournalTitle{Phys. Rev. X}} \textbf{3}, 041019 (2013).

\bibitem{Floess:2018}
D.~Floess and H.~Giessen, \enquote{Nonreciprocal hybrid magnetoplasmonics,}
  {\protect\JournalTitle{Rep. Prog. Phys.}} \textbf{81}, 116401 (2018).

\bibitem{Chekhov:2014}
A.~Chekhov, V.~Krutyanskiy, A.~Shaimanov, A.~Stognij, and T.~Murzina,
  \enquote{Wide tunability of magnetoplasmonic crystals due to excitation of
  multiple waveguide and plasmon modes,} {\protect\JournalTitle{Opt. Express}}
  \textbf{22}, 17762--17768 (2014).

\bibitem{Lukyanchuk:2010}
B.~Luk'yanchuk, N.~Zheludev, S.~Maier, N.~Halas, P.~Nordlander, G.~H., and
  T.~C. Chong, \enquote{The fano resonance in plasmonic nanostructures and
  metamaterials,} {\protect\JournalTitle{Nature Materials}} \textbf{9},
  707--715 (2010).

\bibitem{Kuzmichev:2020}
A.~N. Kuzmichev, D.~A. Sylgacheva, M.~A. Kozhaev, D.~M. Krichevsky, A.~N.
  Shaposhnikov, V.~N. Berzhansky, V.~Berzhansky, F.~Freire-Fern\'andez, H.~J.
  Qin, O.~E. Popova, N.~Keller, S.~van Dijken, A.~I. Chernov, and V.~I.
  Belotelov, \enquote{Influence of the plasmonic nanodisk positions inside a
  magnetic medium on the faraday effect enhancement,}
  {\protect\JournalTitle{Physica Status Solidi (RRL) -- Rapid Research
  Letters}} \textbf{14}, 1900682 (2020).

\bibitem{Krichevsky:2020}
D.~M. Krichevsky, A.~N. Kalish, M.~A. Kozhaev, D.~A. Sylgacheva, A.~N.
  Kuzmichev, S.~A. Dagesyan, V.~G. Achanta, E.~Popova, N.~Keller, and V.~I.
  Belotelov, \enquote{Enhanced magneto-optical faraday effect in
  two-dimensional magnetoplasmonic structures caused by orthogonal plasmonic
  oscillations,} {\protect\JournalTitle{Phys. Rev. B}} \textbf{102}, 144408
  (2020).

\bibitem{MOBook}
A.~Zvezdin and V.~Kotov, \emph{Modern Magnetooptics and Magnetooptical
  Materials} (IOP, 1997).

\bibitem{Kushwaha:1987}
M.~S. Kushwaha and P.~Halevi, \enquote{Magnetoplasma modes in thin films in the
  faraday configuration,} {\protect\JournalTitle{Phys. Rev. B}} \textbf{35},
  3879--3889 (1987).

\bibitem{Belotelov:2014}
V.~I. Belotelov, L.~E. Kreilkamp, A.~N. Kalish, I.~A. Akimov, D.~A. Bykov,
  S.~Kasture, V.~J. Yallapragada, A.~V. Gopal, A.~M. Grishin, S.~I. Khartsev,
  M.~Nur-E-Alam, M.~Vasiliev, L.~L. Doskolovich, D.~R. Yakovlev, K.~Alameh,
  A.~K. Zvezdin, and M.~Bayer, \enquote{Magnetophotonic intensity effects in
  hybrid metal-dielectric structures,} {\protect\JournalTitle{Phys. Rev. B}}
  \textbf{89}, 045118 (2014).

\bibitem{Levy:19}
M.~Levy, O.~Borovkova, C.~Sheidler, B.~Blasiola, D.~Karki, F.~Jomard, M.~A.
  Kozhaev, E.~Popova, N.~Keller, and V.~I. Belotelov, \enquote{Faraday rotation
  in iron garnet films beyond elemental substitutions,}
  {\protect\JournalTitle{Optica}} \textbf{6}, 642--646 (2019).

\bibitem{kalish2014transformation}
A.~Kalish, D.~Ignatyeva, V.~Belotelov, L.~Kreilkamp, I.~Akimov, A.~V. Gopal,
  M.~Bayer, and A.~Sukhorukov, \enquote{Transformation of mode polarization in
  gyrotropic plasmonic waveguides,} {\protect\JournalTitle{Laser Physics}}
  \textbf{24}, 094006 (2014).

\bibitem{BorovkovaMDPI:2022}
O.~Borovkova, S.~Lutsenko, M.~Kozhaev, A.~Kalish, and V.~Belotelov,
  \enquote{Spectrally selective detection of short spin waves in
  magnetoplasmonic nanostructures via the magneto-optical intensity effect,}
  {\protect\JournalTitle{Nanomaterials}} \textbf{12}, 405 (2022).

\bibitem{Borovkova:2020}
O.~Borovkova, H.~Hashim, D.~Ignatyeva, M.~Kozhaev, A.~Kalish, S.~Dagesyan,
  A.~Shaposhnikov, V.~Berzhansky, V.~Achanta, L.~Panina, A.~Zvezdin, and
  V.~Belotelov, \enquote{Magnetoplasmonic structures with broken spatial
  symmetry,} {\protect\JournalTitle{Phys. Rev. B}} \textbf{102}, 081405(R)
  (2020).

\bibitem{Borovkova:2022}
O.~Borovkova, M.~Kozhaev, H.~Hashim, A.~Kolosova, A.~Kalish, S.~Dagesyan,
  A.~Shaposhnikov, V.~Berzhansky, and V.~Belotelov, \enquote{Transverse
  magneto-photonic transmission effect in non-symmetric nanostructures with
  comb-like plasmonic gratings,} {\protect\JournalTitle{Optical Materials
  Express}} \textbf{12}, 573--583 (2022).

\bibitem{Hubert:2008}
A.~Hubert and R.~Sch{\"a}ffer, \enquote{Magnetic domains: the analysis of
  magnetic microstructures,} {\protect\JournalTitle{Springer}}  (2008).

\bibitem{Barturen:2012}
M.~Barturen, B.~R. Salles, P.~Schio, J.~Milano, A.~Butera, S.~Bustingorry,
  C.~Ramos, A.~de~Oliveira, M.~Eddrief, E.~Lacaze, F.~Gendron, V.~H. Etgens, ,
  and M.~Marangolo, \enquote{Crossover to striped magnetic domains in fega
  magnetostrictive thin films,} {\protect\JournalTitle{Appl. Phys. Lett.}}
  \textbf{101}, 092404 (2012).

\bibitem{Dho:2003}
J.~Dho, Y.~Kim, Y.~Hwang, J.~Kim, and N.~Hur, \enquote{Strain-induced magnetic
  stripe domains in la0.7sr0.3mno3 thin films,} {\protect\JournalTitle{Appl.
  Phys. Lett.}} \textbf{82}, 1434 (2003).

\bibitem{Dai:2019}
G.~Dai, X.~Xing, Y.~Shen, and X.~Deng, \enquote{Stress tunable magnetic stripe
  domains in flexible fe81ga19 films,} {\protect\JournalTitle{Journal of
  Physics D: Applied Physics}} \textbf{53}, 055001 (2019).

\bibitem{Lahtinen:2011}
T.~Lahtinen, J.~Tuomi, and S.~van Dijken, \enquote{Pattern transfer and
  electric-field-induced magnetic domain formation in multiferroic
  heterostructures,} {\protect\JournalTitle{Advanced Materials}} \textbf{23},
  3187--3191 (2011).

\bibitem{Satoh:2012}
T.~Satoh, Y.~Terui, R.~Moriya, B.~A. Ivanov, K.~Ando, E.~Saitoh, T.~Shimura,
  and K.~Kuroda, \enquote{Directional control of spin-wave emission by
  spatially shaped light,} {\protect\JournalTitle{Nature Photonics}}
  \textbf{6}, 662--666 (2012).

\bibitem{Nikitov:2015}
S.~Nikitov, D.~Kalyabin, I.~Lisenkov, A.~Slavin, Y.~Barabanenkov, S.~Osokin,
  A.~Sadovnikov, E.~Beginin, M.~Morozova, Y.~Sharaevsky, Y.~Filimonov,
  Y.~Khivintsev, S.~Vysotsky, V.~Sakharov, and E.~Pavlov, \enquote{Magnonics: a
  new research area in spintronics and spin wave electronics,}
  {\protect\JournalTitle{Phys. Usp.}} \textbf{58}, 1002--1028 (2015).

\bibitem{Kimel:2005}
A.~V. Kimel, A.~Kirilyuk, P.~A. Usachev, R.~V. Pisarev, A.~M. Balbashov, and
  T.~Rasing, \enquote{Ultrafast non-thermal control of magnetization by
  instantaneous photomagnetic pulses,} {\protect\JournalTitle{Nature}}
  \textbf{435}, 655--657 (2005).

\bibitem{Kirilyuk:2010}
A.~Kirilyuk, A.~Kimel, and T.~Rasing, \enquote{Ultrafast optical manipulation
  of magnetic order,} {\protect\JournalTitle{Reviews of Modern Physics}}
  \textbf{82}, 2731 (2010).

\bibitem{Kozhaev:2018}
M.~Kozhaev, A.~Chernov, D.~Sylgacheva, A.~Shaposhnikov, A.~Prokopov,
  V.~Berzhansky, A.~Zvezdin, and V.~I. Belotelov, \enquote{Giant peak of the
  inverse faraday effect in the band gap of magnetophotonic microcavity,}
  {\protect\JournalTitle{Scientific Reports}} \textbf{8}, 11435 (2018).

\bibitem{Savochkin:2017}
I.~Savochkin, M.~Jäckl, V.~Belotelov, I.~Akimov, M.~Kozhaev, D.~Sylgacheva,
  A.~Chernov, A.~Shaposhnikov, A.~Prokopov, V.~Berzhansky, D.~Yakovlev,
  A.~Zvezdin, and M.~Bayer, \enquote{Generation of spin waves by a train of
  fs-laser pulses: a novel approach for tuning magnon wavelength,}
  {\protect\JournalTitle{Sci. Rep.}} \textbf{7}, 5668 (2017).

\bibitem{Popova:2012}
E.~Popova, L.~Magdenko, H.~Niedoba, M.~Deb, B.~Dagens, B.~Berini,
  M.~Vanwolleghem, C.~Vilar, F.~Gendron, A.~Fouchet, J.~Scola, Y.~Dumont,
  M.~Guyot, and N.~Keller, \enquote{Magnetic properties of the magnetophotonic
  crystal based on bismuth iron garnet,} {\protect\JournalTitle{J. Appl.
  Phys.}} \textbf{112}, 093910 (2012).

\bibitem{Deb:2012}
M.~Deb, E.~Popova, A.~Fouchet, and N.~Keller, \enquote{Magneto-optical faraday
  spectroscopy of completely bismuth-substituted bi3fe5o12 garnet thin films,}
  {\protect\JournalTitle{Journal of Physics D: Applied Physics}} \textbf{45},
  455001 (2012).

\bibitem{Moharam:1995}
M.~Moharam, E.~Grann, D.~Pommet, and T.~Gaylord, \enquote{Formulation for
  stable and efficient implementation of the rigorous coupled-wave analysis of
  binary gratings,} {\protect\JournalTitle{J. Opt. Soc. Am.}} \textbf{12}, 1068
  (1995).

\bibitem{Li:2003}
L.~Li, \enquote{Fourier modal method for crossed anisotropic gratings with
  arbitrary permittivity and permeability tensors,} {\protect\JournalTitle{J.
  Opt. A}} \textbf{5}, 345 (2003).

\end{thebibliography}

\end{document}